\documentstyle[sprocl]{article}

\input{psfig.sty}

\def\Journal#1#2#3#4{{#1} {\bf #2}, #3 (#4)}

\def\NPB{{\em Nucl. Phys.} B}

\def\PRD{{\em Phys. Rev.} D}

\def\be{\begin{equation}}
\def\ee{\end{equation}}
\def\bea{\begin{eqnarray}}
\def\eea{\end{eqnarray}}

\begin{document}

\title{NUCLEON SPIN STRUCTURE AND LARGE $p_T$ PROCESSES AT $pp$
COLLIDERS \footnote{Talk presented at SPIN98 symposium, Protvino, Russia, 
September 8-12, 1998, in collab. with G. P. Ramsey} }

\author{L. E. Gordon}

\address{Jefferson Lab, Newport News and
Hampton University, Hampton, VA 23606, USA} 


\maketitle
\abstracts{ 
QCD motivated polarized parton distributions, evolved directly in $x$-space,
are used to predict rates for prompt photon production at RHIC 
centre of mass energies. Various scenarios for the polarized 
gluon distributions are considered and compared, and the possibility of 
using large $p_T$ processes in polarized $pp$ collision experiments to 
choose between them is analyzed.  
}

\section{Introduction}

In a recent paper \cite{ggr} we proposed three new parametrizations for
the spin dependent parton distributions for the proton. Our
distributions, which are available in both leading order (LO) and
next-to-leading order (NLO), were evolved directly in $x$-space. In
\cite{ggr} we presented details of the models used to obtain the input 
distributions and compared the evolved structure functions with the available 
data. 

All currently available data on the nucleon spin structure come from 
deep inelastic scattering experiments, and therefore do not contain
direct information on individual parton distributions such as the gluon,
$\Delta G(x,Q^2)$ or strange sea, $\Delta s(x,Q^2)$, for example. It is
expected that such detailed information will come from other experiments
such as those proposed at the Relativistic  Heavy Ion Collider (RHIC) at
Brookhaven or HERA-$\vec{N}$ in Hamburg.   

One of the main programs at RHIC will be to determine the size and the 
shape of the polarized parton distributions which, at the moment, suffer from 
significant model dependence, especially in the gluon contribution,
$\Delta G$, but also the sea quark distributions.

In this paper we use our evolved parton
distributions to predict cross sections for direct photon 
production at RHIC cms energies. Details of the evolution of the parton
distributions and predictions for prompt photon production can be found
in \cite{gr}.

The prompt photon cross section promises to be one of the most useful for
measuring $\Delta G$ at RHIC and HERA-$\vec{\rm N}$ since it is dominated by
the subprocess $qg\rightarrow \gamma q$.
Contributions to the prompt photon cross section are usually separated
into two classes in both LO and NLO. There are the so-called direct
processes where the photon is produced 
directly in the hard scattering. In addition there are the fragmentation
contributions where the photon is produced via bremsstrahlung off a
final state quark or gluon. 

\section{Results}

In this paper we present results for the inclusive prompt photon cross section
at RHIC energies
without taking any possible isolation cuts into account. As shown in
\cite{leg}, isolation cuts do not have a significant effect on
the asymmetries which are the quantities in which we are mainly
interested here. 
 
Figure 1 shows our various parametrizations of the polarized gluon
distributions at $Q^2=100$ GeV$^2$. The GSA \cite{gs} parametrization is
included for comparison. 
Our polarized gluon distributions are smaller than GSA, which affects the
ratio of quark to gluon contributions at small $x$. This in turn will
modify the relative contributions to direct $\gamma$ production. 
(See figure 3).

Figure 2 shows the asymmetries, $A_{LL}$, as predicted using the various
parametrizations at $\sqrt{s}=200$ GeV. 
The asymmetries differ significantly confirming sensitivity to $\Delta
G $. They differ from GSA most significantly at the largest $p_T$ values 
($>40 \;GeV$), where RHIC is at the
$p_T$ limit. However, measurement of this asymmetry should be able to
distinguish between small and large polarized gluon distributions at large
$p_T$. 

Figure 3 confirms that the $qg$ scattering process dominates the direct
$\gamma$ cross section at the $p_T$ values where the cross section is
largest. The second most important process $q\bar{q}$ scattering only
becomes significant at very large $p_T$ values.

\section{Conclusions}
 
The new NLO GGR parametrizations of the polarized parton distributions
were used to predict asymmetries for direct photon production at RHIC
energies. The three scenarios considered give different predictions
which will be distinguishable with reasonably good statistics in the
measurements.

\begin{figure}[t] 
\hspace{2.5 cm}
\psfig{file=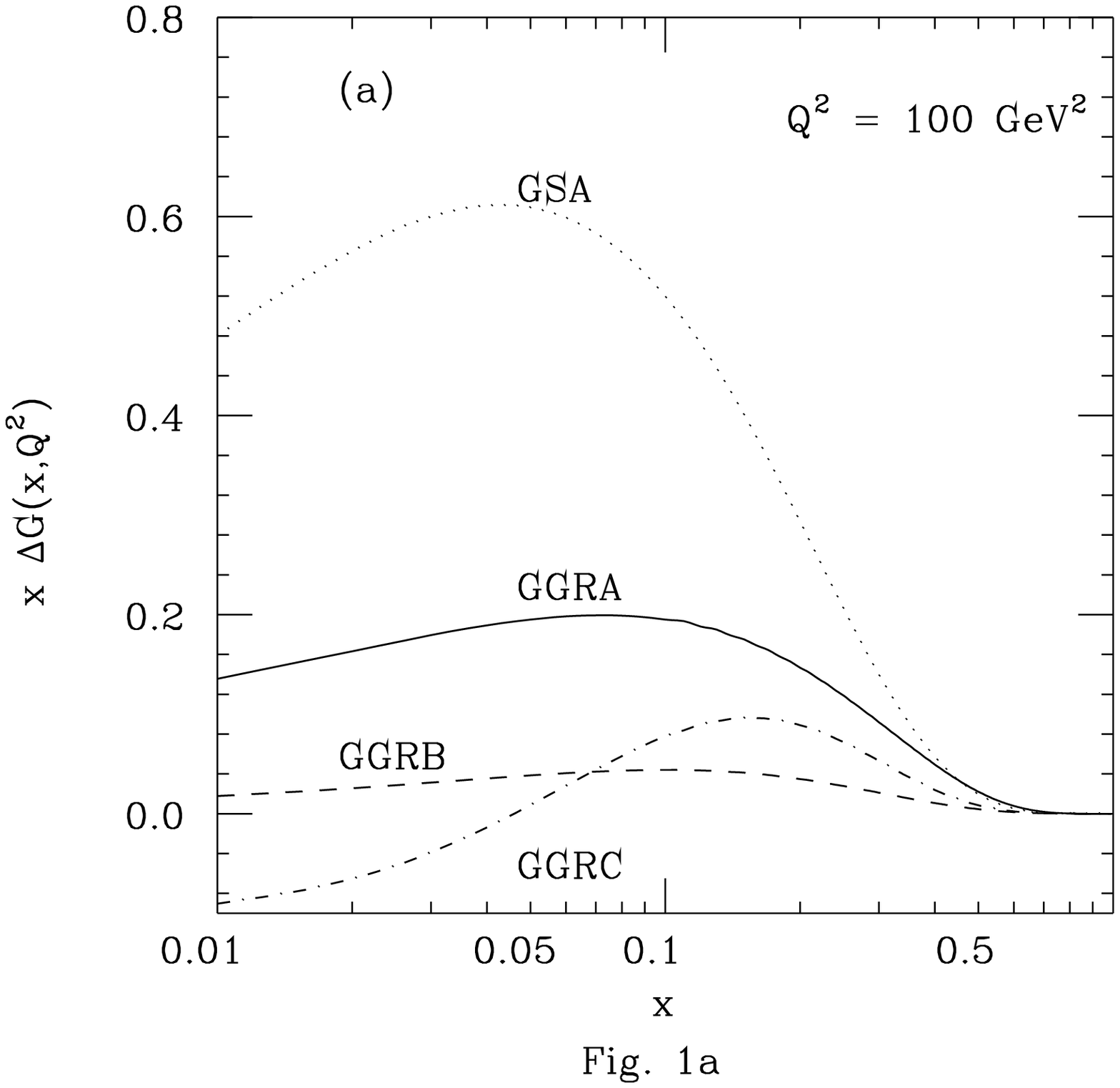,height=4.5cm,width=9 cm} 
\caption{The GGR polarized gluon
distributions compared to GSA. } 
\end{figure}
\begin{figure}[t] 
\hspace{2.5cm} 
\psfig{file=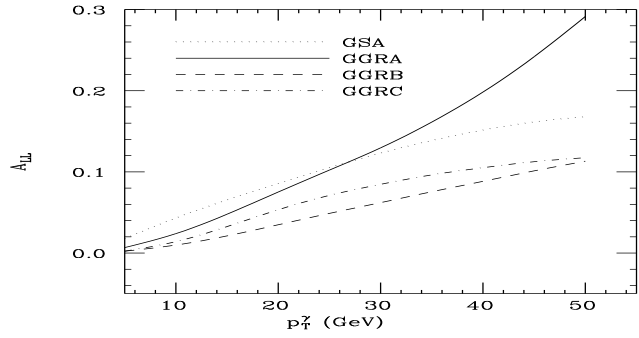,height=4.5cm,width=9 cm}
\caption{Asymmetries for direct $\gamma$ production.}  
\end{figure}
\begin{figure}[t] 
\hspace{2.5 cm}
\psfig{file=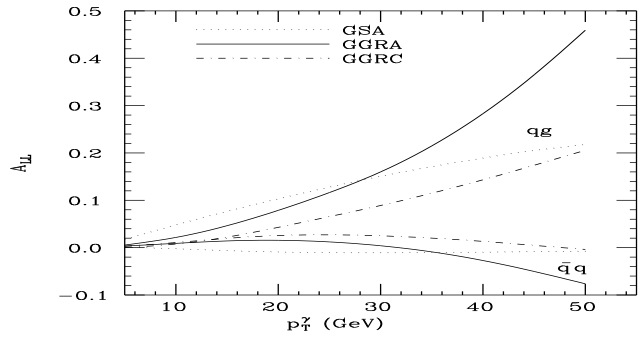,height=4.5cm,width=9 cm}  
\caption{Asymmetries showing $qg$ and $q\bar{q}$ contributions.}  
\end{figure}

\end{document}